\documentclass[aps,prl,superscriptaddress,amsfonts,11pt]{revtex4-2}

\usepackage{amsmath}
\usepackage{multirow}
\usepackage{amsmath,amssymb,graphicx}
\usepackage[colorlinks=true, pdfstartview=FitV,
linkcolor=blue, citecolor=blue, urlcolor=blue]{hyperref}
\usepackage[utf8]{inputenc}
\usepackage{color}
\usepackage{lmodern}
\usepackage{float}
\usepackage[normalem]{ulem}
\usepackage{cancel}

\makeatletter
\def\btt#1{\texttt{\@backslashchar#1}}%
\DeclareRobustCommand\bblash{\btt{\@backslashchar}}%
\makeatother

\definecolor{fj_color}{cmyk}{1, 0.3, 0, 0}
\definecolor{cwh_color}{cmyk}{0, 0.8, 0.8, 0}
\definecolor{hr_color}{cmyk}{0.5, 0.8, 0.0,0}

\newcommand{\BKFA}{Ba\textsubscript{1-x}K\textsubscript{x}Fe\textsubscript{2}As\textsubscript{2}}

\begin{document}

\preprint{PREPRINT (\today)}

\title{Calorimetric evidence for two phase transitions in \BKFA{} with fermion pairing and quadrupling states
}

\author{Ilya Shipulin}
\email{This authors equally contributed to the current work}
\affiliation{Institute for Metallic Materials,  Leibniz-IFW Dresden, D-01069, Dresden, Germany}

\author{Nadia Stegani}
\email{This authors equally contributed to the current work}
\affiliation{University of Genoa, Via Dodecaneso 33, 16146 Genoa, Italy}
\affiliation{Consiglio Nazionale delle Ricerche (CNR)-SPIN, Corso Perrone 24, 16152 Genova, Italy}

\author{Ilaria Maccari}
\affiliation{Department of  Physics, Stockholm University, Stockholm SE-10691, Sweden}

\author{Kunihiro Kihou} 
\affiliation{National Institute of Advanced Industrial Science and Technology (AIST), Tsukuba, Ibaraki 305-8568, Japan}

\author{Chul-Ho Lee} 
\affiliation{National Institute of Advanced Industrial Science and Technology (AIST), Tsukuba, Ibaraki 305-8568, Japan}
\author{Yongwei Li}
\affiliation{Tsung-Dao Lee Institute, Shanghai Jiao Tong University, Pudong, Shanghai, China}

\author{Ruben H\"uhne}
\affiliation{Institute for Metallic Materials,  Leibniz-IFW Dresden, D-01069, Dresden, Germany}

\author{Hans-Henning Klauss}
  \affiliation{Institute for Solid State and Materials Physics, Technische Universit\"{a}t Dresden, D-01069 Dresden, Germany}

\author{Marina Putti}
\affiliation{University of Genoa, Via Dodecaneso 33, 16146 Genoa, Italy}
\affiliation{Consiglio Nazionale delle Ricerche (CNR)-SPIN, Corso Perrone 24, 16152 Genova, Italy}

\author{Federico Caglieris}
\email{federico.caglieris@spin.cnr.it}
\affiliation{University of Genoa, Via Dodecaneso 33, 16146 Genoa, Italy}
\affiliation{Consiglio Nazionale delle Ricerche (CNR)-SPIN, Corso Perrone 24, 16152 Genova, Italy}
\affiliation{Institute for Metallic Materials,  Leibniz-IFW Dresden, D-01069, Dresden, Germany}

\author{Egor Babaev}
\email{babaev.egor@gmail.com}
\affiliation{Department of Physics, KTH Royal Institute of Technology, SE-106 91 Stockholm, Sweden}

\author{Vadim Grinenko}
\email{vadim.a.grinenko@gmail.com}
\affiliation{Tsung-Dao Lee Institute, Shanghai Jiao Tong University, Pudong, Shanghai, China}

\begin{abstract}

Theoretically, materials that break multiple symmetries allow, under certain conditions, the formation of four-fermion condensates above the superconducting critical temperature. Such states can be stabilized by phase fluctuations. Recently a fermionic quadrupling condensate that breaks the $Z_2$ time-reversal symmetry was reported in \BKFA{} [V. Grinenko et al., Nat. Phys. 17, 1254 (2021)].  Evidence for the new state of matter comes from muon-spin rotation, transport, thermoelectric, and ultrasound experiments. Observing a specific heat anomaly is a very important signature of a transition to a new state of matter. However, a  fluctuation-induced specific heat singularity is usually very challenging to resolve from a background of other contributions.
Here, we report on detecting two anomalies in the specific heat of \BKFA{} at zero magnetic field. The anomaly at the higher temperature is accompanied by the appearance of a spontaneous Nernst effect, indicating broken time-reversal ($Z_2$) symmetry. The second anomaly at the lower temperature coincides with the transition to a zero resistance state, indicating superconductivity breaking the $U(1)$ gauge symmetry. Our data provide 
calorimetric evidence for the $Z_2$ phase formation above the superconducting phase transition.

\end{abstract}

\maketitle

\section{Introduction}
The formation of bound states of fermions can lead to
new states of matter: symmetry-breaking condensates.
The most investigated case is the condensate of paired electrons (Cooper pairs). It results in a new state of matter: superconductivity, characterized by a spontaneously broken U(1) gauge symmetry ~\cite{Bardeen1957a, Bardeen1957b, Ginzburg1950}. 
A complex order parameter $\Delta$ describing the simplest single-band spin-singlet superconducting state is associated with non-vanishing averages of two-fermion annihilation operators $<c_{\uparrow } c_{\downarrow}>$. It describes the flow of Cooper pairs, each carrying twice the electron charge ``e".
Bound states of 4, 6 etc. electrons would also be bosons, but 
within the standard Bardeen–Cooper–Schrieffer (BCS) theory, such condensates do not form. 
Four-fermion electronic condensates can appear via a fluctuations-driven mechanism if the system breaks multiple symmetries \cite{babaev2004superconductor,Babaev2004phase,Bojesen2013,berg2009charge,agterberg2008dislocations,Herland2010}. 
Among four-fermion orders, there is a counterpart of Cooper pair superconductivity: a charge-4e superconducting order parameter can be constructed involving nonzero averages of the kind
 $<c_{\uparrow \rm i} c_{\downarrow \rm i}
 c_{\uparrow \rm j} c_{\downarrow \rm j}>$, where $ \rm i$ and $ \rm j$ are component indices.
  Although that state also breaks $U(1)$ local symmetry and it is a superconductor, it is different in several important aspects from charge-2e superconductors: the coupling constant of the order parameter to  the electromagnetic field is proportional to $4e$ leading to several different effects such as half-quanta vortex excitations. These properties are used in current attempts to find such states in experiments     \cite{ge2022discovery}.
  A bosonic counterpart of such states was also discussed in the context of an ultracold atomic mixture close to the Mott insulating state \cite{kuklov2004superfluid}.

 Besides the superconducting order, one can construct a great diversity of 
 order parameters
 out of four fermionic operators. The variety of the potential new states described by such order parameters is much greater than possible orders arising from fermionic pairs.  
An especially interesting possibility
 is associated with fermion quadrupling condensates  forming above the superconducting phase transition, leading to condensates with principally different properties than superconductors.
 One such possible  state appears when a fermion quadrupling condensate results in a
 Broken Time-Reversal Symmetry (BTRS)
 ~\cite{Bojesen2013,Bojesen2014a,Grinenko2021state,Garaud2022,Maccari2022effects,maccari2022possible}.
At low temperatures, such a system is a multicomponent superconductor that breaks time-reversal symmetry and can be
 described by several complex fields $\Delta_{\rm i}$, where $\rm i$ is a component or band index \cite{Maiti2013,Boeker2017,Carlstrom2011b,Stanev2010}.
 Since applying the time-reversal operation twice returns the system to its original state, such a condensate breaks an additional two-fold (i.e. $Z_2$) symmetry. A new state of matter forms under temperature increase  \cite{Grinenko2021state}. In this state, the averages of the pairing order parameters in each band are zero $\langle\Delta_{  \rm i}\rangle=0$, but a nonvanishing composite order parameter $\Delta_{ \rm 4f} \propto \langle\Delta^*_{ \rm i}\Delta_{\rm  j}\rangle\ne 0$ appears \cite{Grinenko2021state,Maccari2022effects,Bojesen2013,Bojesen2014a}. 
 This order parameter is of fourth order in fermionic fields  $<c_{\uparrow \rm i} c_{\downarrow \rm i}
 c^\dagger_{\uparrow \rm j} c^\dagger_{\downarrow \rm j}>$.
The state preserves the local $U(1)$ symmetry and hence it is resistive to dc current. Instead, it breaks $Z_2$ time-reversal symmetry resulting in dissipationless local counterflows of charges between $\rm i$ and $ \rm j$ components.  These currents produce spontaneous magnetic fields around certain kinds of inhomogeneities and topological defects \cite{Grinenko2021state,Garaud2022}.
Many novel properties of this state can be described by an effective model, which is different from the Ginzburg-Landau effective models of superconductors and the Gross-Pitaevskii effective models of superfluids, but is rather related to the Skyrme model \cite{Garaud2022}. The weak spontaneous magnetic fields appearing at $T_{\rm c}^{\rm Z2}$, above the superconducting critical temperature $T_{\rm c}$, and detected by $\mu$SR in \BKFA{} are consistent with   the appearance of a spontaneous Nernst effect  at the same temperature \cite{Grinenko2021state}. This, and other
  experimental data in ref.~\cite{Grinenko2021state} provides the  evidence for a 4-fermionic (quartic) state, which exists in a range of temperatures above $T_{\rm c}$. 

A transition from a normal state to a new state, such as the  quartic state, 
should result in a specific heat anomaly at $T_{\rm c}^{\rm Z2}$. 
A second anomaly at $T_{\rm c}$ should follow at a lower temperature.
An example of these anomalies from Monte-Carlo simulations of a multiband model is shown in the theoretical analysis section below. However, the predicted two anomalies are expected to be difficult to detect experimentally since the quartic phase is a fluctuation-induced effect, in which phase fluctuations cause a tiny contribution on top of a background due to other degrees of freedom. The phase-fluctuations contribution, in addition, 
can be washed out by inhomogeneities since $T_{\rm c}^{\rm Z2}$ strongly depends on doping \cite{Grinenko2020}. 
  Such double anomalies in zero external fields were not observed in experiments so far. Here, we investigate new samples of \BKFA{} with slightly different doping compared to the one studied in ref.~\cite{Grinenko2021state} and report the observation of two anomalies in the zero-field specific heat.
Importantly, these anomalies are consistent with spontaneous Nernst and electrical resistivity data, signalling two separate $Z_2$ and $U(1)$ phase transitions. These data provide calorimetric evidence for a phase transition above $T_{\rm c}$ associated with the formation of the quartic state.

\section{Experimental Results}

\begin{figure}
	\centering
	\includegraphics[width=0.95\linewidth]{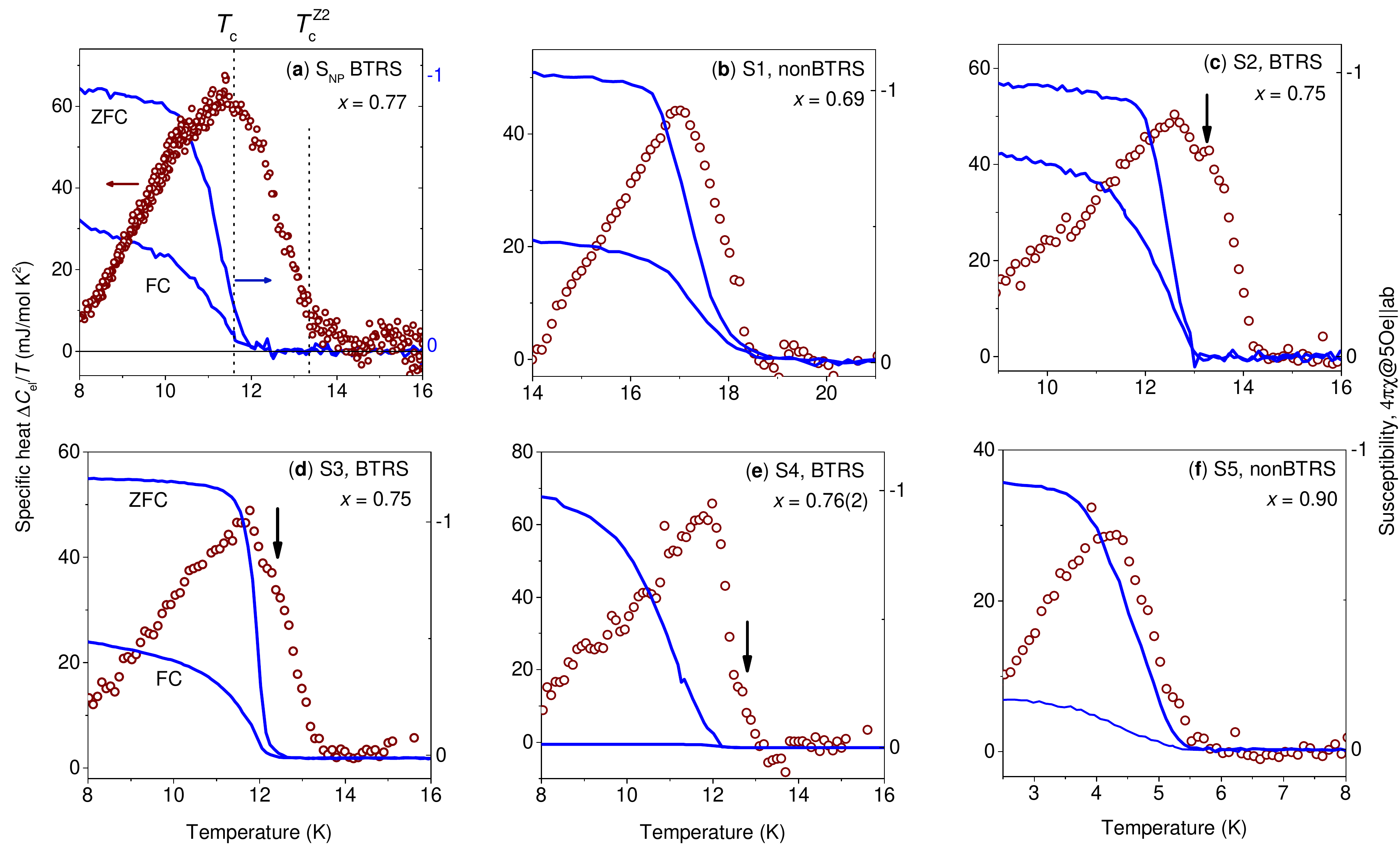}
	\caption{\textbf{Split transitions in hole overdoped \BKFA{}} Temperature dependence of the zero-field specific heat $\Delta C_{\rm el}/T$ and the static magnetic susceptibility measured in $B\parallel ab$=0.5 mT. Panel {\bf (a)} shows the data for the sample from ref.~\cite{Grinenko2021state} with the quartic metal phase and doping level $x$ = 0.77. $T_{\rm c}$ is defined by zero resistance and $T_{\rm c}^{\rm Z2}$ is shown according to the onset of the spontaneous magnetic fields and spontaneous Nernst effect.   A distinct specific heat anomaly at $T_{\rm c}^{\rm Z2}$ was not resolved within error-bars of the measurements. Panels {\bf (b-f)} show the data for the samples studied in this work. The panels {\bf (b)} and {\bf (f)} correspond to reference samples with doping levels  $x\approx$ 0.69   and $x\approx$ 0.90, respectively. These two samples have  conventional superconductivity and the outside the range of doping, where BTRS occurs \cite{Grinenko2017bkfa,Grinenko2020}. These two samples show the conventional picture, where the onset of the specific heat is consistent  with the onset of a superconducting response in the susceptibility, i.e. indistinguishable from the standard mean-field behavior of a superconductor with a single phase transition at $T_{\rm c}$. The panels {\bf (c-e)} are for the samples with BTRS state and doping level $x\approx$ 0.75. These sample show two anomalies in specific heat. The second anomaly, highlighted by black arrows,  appears at the $Z_2$ transition (see Fig.~\ref{fig3}).}

	\label{fig2}
\end{figure}

\begin{figure}
	\centering
	\includegraphics[width=0.8\linewidth]{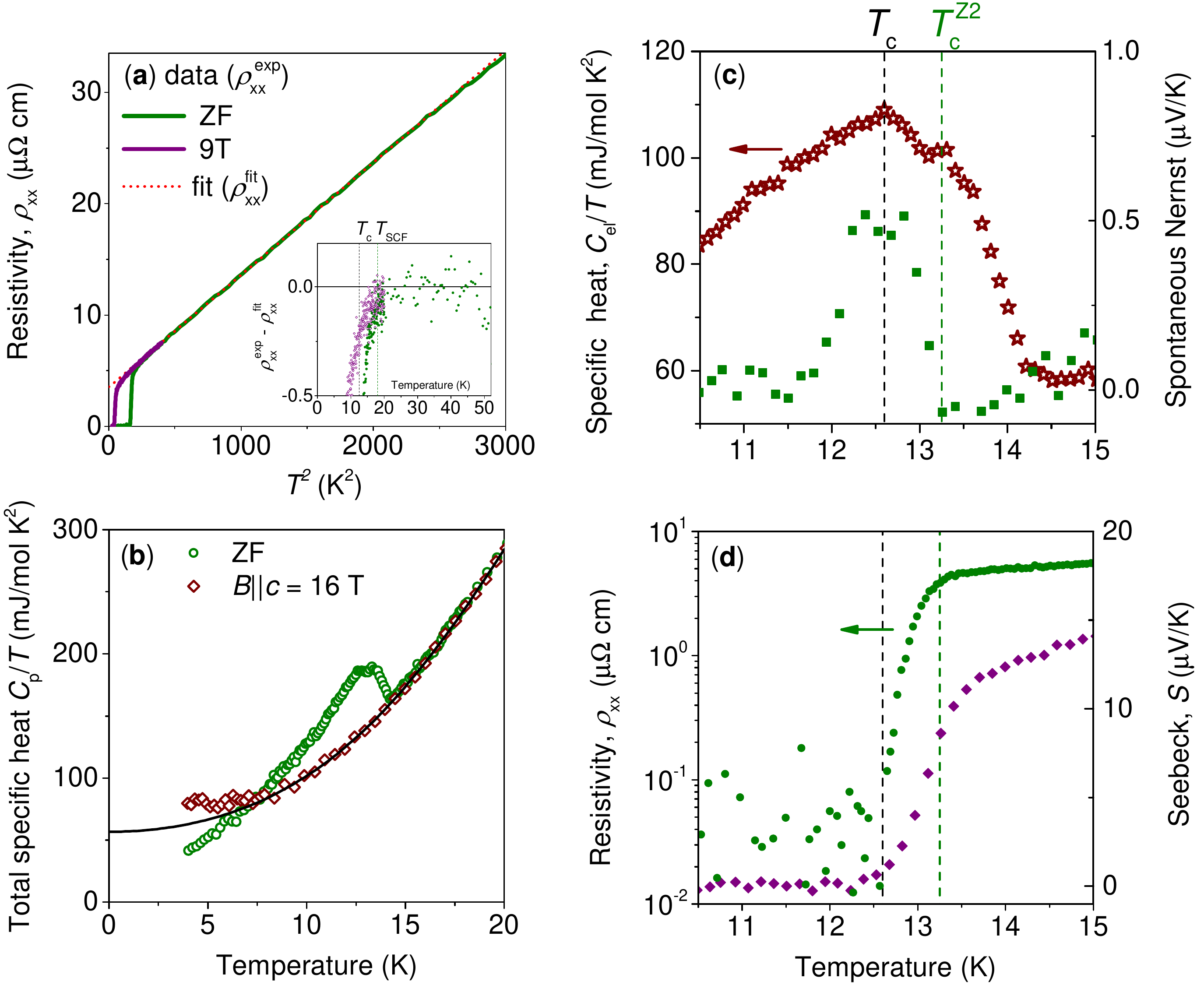}
	\caption{\textbf{Two phase transitions in \BKFA{}} The figure shows the experimental data for $S2$ sample. \textbf{(a)} The longitudinal electrical resistivity vs. temperature squared measured in zero and 9 T field applied along the crystallographic $c$-axis. The inset shows the temperature dependence of the difference between the fit curves and the experimental data. The resistivity deviates from the normal-state behaviour at $T_{\rm SCF}$ associated with superconducting fluctuations. \textbf{(b)} Temperature dependence of the specific heat measured in zero and 16 T field applied along the crystallographic $c$-axis. The solid line is a fit to approximate the normal state behavior. A two-step anomaly is seen in the zero field specific heat data. \textbf{(c)} (left axis) Temperature dependence of the difference between the zero field specific heat experimental data and the phonon contribution defined using the 16 T experimental data from panel (b) compared with (right axis) temperature dependence of the zero field spontaneous Nernst effect. The appearance of the spontaneous Nernst effect signals a spontaneous breaking of time-reversal symmetry at $T_{\rm c}^{\rm Z2}$. It  coincides with the high-temperature anomaly in the specific heat. \textbf{(d)}  Temperature dependence of the zero field electrical resistivity (left axis) and the Seebeck effect. The specific heat in panel (c) shows the  second anomaly appearing at a lower temperature corresponding to the superconducting $T_{\rm c}$.}
	\label{fig3}
\end{figure}

The temperature dependence of the specific heat and the magnetic susceptibility measured on the sample ($S_{\rm NP}$) from Ref.~\cite{Grinenko2021state} is shown in panel (a) of Fig.~\ref{fig2}. There is a significant splitting between the onset temperatures for the specific-heat anomaly and the diamagnetic susceptibility.
As discussed in Ref.~\cite{Grinenko2021state}, this splitting is related to the precursor formation of electronic bound states and eventually to the $Z_2$ phase transition above the $T_{\rm c}$. However, the expected distinct anomaly in the specific heat at $T_{\rm c}^{\rm Z2}$ was not resolved in samples investigated in ref.~\cite{Grinenko2021state}.
In this work, we performed systematic specific heat measurements of several new samples ($S2 - S4$) with a doping level of $x\approx 0.75$, close to $x= 0.77$ for $S_{\rm NP}$ sample.
According to the previous studies \cite{Grinenko2017bkfa,Grinenko2020}, this doping level is within the range where the superconducting state breaks time-reversal symmetry. In addition, we had two reference samples ($S1$, and $S5$) that do not break time-reversal symmetry. The specific-heat data $\Delta C_{\rm el}/T$ are summarized in Fig.~\ref{fig2}. The raw data are given in extended data Fig.~ED1 for samples $S1$, $S3$-$S5$, and in Fig.~\ref{fig3}b for sample $S2$.

The reference samples (panels (b) and (f) in Fig.~\ref{fig2}) show conventional behaviour with a single phase transition at $T_{\rm c}$ expected for standard superconductors: the appearance of non-dissipative currents observed in the susceptibility data match 
with the onset of the specific-heat anomaly. The samples with the BTRS superconducting state (panels (c-e)) show a qualitatively different behaviour, which is similar to the previously investigated  $S_{\rm NP}$ sample shown on panel (a). In contrast to sample $S_{\rm NP}$, the samples investigated in this work show a step-like anomaly above $T_{\rm c}$. These anomalies cannot be attributed to a superconducting phase with higher $T_{\rm c}$ since no superconducting signal is observed in the magnetic susceptibility at the corresponding temperature (right axis in Fig.~\ref{fig2}). For a detailed analysis of a possible inhomogeneity effect on the susceptibility and specific heat, see \cite{Grinenko2021state}.

To investigate whether this anomaly can be associated with the $Z_2$ time-reversal-symmetry-breaking phase transition, we performed more detailed investigations on sample $S2$, which has the most pronounced double anomalies in the specific heat. The data are summarized in Fig.~\ref{fig3} and Fig.~ED2. Fig.~\ref{fig3}a shows the electrical resistivity plotted versus squared temperature. The resistivity follows a conventional Fermi liquid behaviour in the normal state with the residual resistivity ratio $RRR\approx70$. This indicates high sample quality and the absence of proximity to magnetic critical points. The inset in panel (a) shows the difference between the experimental data and the $T^2$ fitting curve. It is seen that resistivity deviates from $T^2$-behavior below $T_{\rm SCF}\approx$ 18 K. $T_{\rm SCF}$ is associated with the onset of detectable effects of superconducting fluctuations  ~\cite{Grinenko2021state}. Note that, compared to about 10 K for a sample $S_{\rm NP}$, here  the temperature difference between $T_{\rm SCF}$ and $T_{\rm c}\approx$ 12.6 K is smaller.  

The specific heat data measured in zero and applied magnetic field $B\parallel c$ = 16 T are shown in panel (b).  The magnetic field doesn't suppress the specific heat anomaly completely. Therefore, to subtract the phonon contribution from the zero-field specific heat, we fit the in-field specific heat above 10 K and use the fitting curve to represent the  values from the phonon background. Details of the fitting procedure can be found in Refs.~\cite{Grinenko2017bkfa,Grinenko2020}. The result is shown in panel (c), left axis, and compared with the temperature dependence of the zero-field spontaneous Nernst effect, right axis. The onset of the spontaneous Nernst signal gives the critical temperature of the BTRS transition at $T_{\rm c}^{\rm Z2}\approx$13.25 K, indicating the formation of the 4-fermion ordered state characterized by intercomponent phase locking  arising at a significantly lower temperature than the crossover associated with local superconducting fluctuations $T_{\rm SCF}$. That is consistent with the theoretical expectations that the relative phase ordering of the four-fermion order parameter $\langle\Delta^*_{\rm i}\Delta_{\rm j}\rangle\ne 0$ should occur below the onset of incoherent fluctuations. Remarkably, as shown in panel (c), the $T_{\rm c}^{\rm Z2}$ transition temperature coincides with the high-temperature anomaly in the specific heat.

Next, our data shows that $T_{\rm c}^{\rm Z2}$ splits from the superconducting critical temperature $T_{\rm c}$. 
For this purpose, the temperature dependence of the zero-field electrical resistivity (left axis) is compared with the Seebeck effect (right axis) in panel (d) of Fig.~\ref{fig3}. Both quantities are signalling superconducting phase transitions 
at $T_{\rm c}\approx$12.6 K defined by the temperature at which the resistivity and the Seebeck effect are zero. This temperature is lower than $T_{\rm c}^{\rm Z2}\approx$13.25 K. Notably, $T_{\rm c}$ coincides with the lower anomaly  (sharp maximum) in the specific heat, indicating that the appearance of zero resistance is caused by the appearance of a finite superconducting order parameter \cite{Grinenko2021state}. Note that the double-step anomaly in the zero-field specific heat is well visible in the raw data shown in panel (b). These observations allow us to conclude that our data provide calorimetric evidence for a $Z_2$ phase transition above $T_{\rm c}$.

\section{Theoretical analysis}
\begin{figure}[tbh]
	\centering
	\includegraphics[width=0.65
\linewidth]{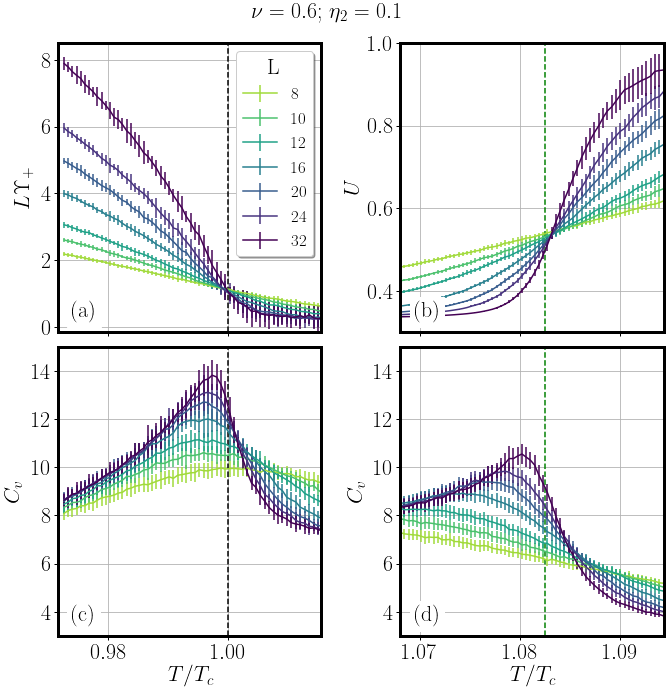}
	\caption{{\bf An example of double specific heat anomalies. (a,b)} 
	The results from Monte-Carlo computations in the simplest London  model with electron quadrupling BTRS state from Refs \cite{Grinenko2021state,Maccari2022effects} (model description can found in the Supplementary information) {\bf(a)} Helicity modulus  $\Upsilon_+$ as a function of the temperature, signalling the superconducting phase transition at $T_{\rm c}$. 
	{\bf(b)} Binder cumulant $U$ versus $T/T_{\rm c}$ signalling the breakdown of the time reversal symmetry at $T_{\rm c}^{\rm Z2}$.  {\bf(c,d)} The calculated specific heat as function of $T/T_{\rm c}$, for temperatures close to the $T_{\rm c}$ {\bf(c)} and $T_{\rm c}^{\rm Z2}$ {\bf(d)} transition temperatures in zero magnetic field. The black and the green dashed lines indicate respectively the $T_{\rm c}$ and the $T_{\rm c}^{\rm Z2}$ calculated in the thermodynamic limit as explained in the Supplementary information.  
	The model gives the contribution to the specific heat from the phase fluctuations, only. In a real system such as \BKFA{}, this contribution should be superimposed with the non-singular  contribution from pair-breaking effects.
 }
	\label{fig1}
\end{figure}

To demonstrate the two singularities in the specific heat we use the simplest phenomenological two-component
 free-energy functional  that yields a similar phase diagram (for  detail, see ~\cite{Maccari2022effects,Garaud2017,Garaud2022}) 
 We use a simple London model with two phases $\phi_{1,2}$
originating from two complex fields $\Psi_{\alpha}= |\Psi_{\alpha}| e^{i \phi_{\alpha}}$. The existence of a fermion quadrupling phase in that model was
previously reported in \cite{Grinenko2021state,Maccari2022effects} without discussing the specific heat.

The model
\begin{equation}
        f= \frac{1}{2} \sum_{\alpha=1,2}  \left( {\mathbf{\nabla}} \phi_{\alpha} \right)^2  -\nu \left({\mathbf{\nabla}} \phi_1  \cdot {\mathbf{\nabla}}\phi_2\right)+ \eta_2 \cos[2(\phi_1  -\phi_2  )].
\label{f_2component}
\end{equation}
accounts for two different inter-component interactions: the second-order biquadratic Josephson interaction with a coupling constant $\eta_2>0$, and the mixed-gradient term with a coupling constant $\nu>0$.
  The presence of the biquadratic term reduces the total symmetry of the model to $U(1)\times Z_2$, where the $Z_2$ time-reversal symmetry is associated with the two equivalent minima for the phase-difference mode, i.e. $\phi_{1,2}= \phi_1 - \phi_2 = \pm \pi/2$,  spontaneously broken at low temperatures. 
This represents one of the simplest models with a BTRS fermion quadrupling phase in zero external magnetic field  \cite{Maccari2022effects}.
Here we consider the model situation where we neglect the coupling to an electromagnetic field. Finite values of the magnetic field penetration length increase the size of the fermion-quadrupling phase \cite{Maccari2022effects}.
Here, we discuss the specific heat of the system in the presence of a finite splitting between the two critical temperatures of the order of 10 $\%$ according to experimental observations. We note that in multicomponent models, when transition splitting is small, a transition could become weakly first order \cite{Kuklov2006,Svistunov2015}, which would alter the specific heat picture.
As shown in Fig.\ref{fig1}, our Monte-Carlo (MC) simulations reveal that the specific heat $C_v$ develops two anomalies in the proximity of the two critical temperatures $T_{\rm c}$ and $T_{\rm c}^{\rm Z2}$, respectively. We note that the model we consider is a phase-only model that only accounts for the phase-fluctuations contribution to the specific heat. In a real system, such as \BKFA{}, this will represent small additional  contributions superimposed  with the non-singular smooth contributions from pair-breaking effects. Note that Monte-Carlo calculations are done on finite-size systems, and real superconductors always have some degree of inhomogeneity. Hence for both transitions, instead of divergence of specific heat, one observes anomalies in the form of finite ``bumps". The obtained ratio between amplitudes of the anomalies is relatively similar to  the experimental observations for similar splitting between $T_{\rm c}$ and $T_{\rm c}^{\rm Z2}$ transition temperatures.

\section{Discussion}
A specific heat singularity occurs when a system transitions into another state. The electron quadrupling phases of the kind that we discuss in this paper are stabilized by fluctuations. The experimental observation of a specific heat anomaly associated with phase fluctuations is usually difficult in superconducting systems due to the relatively small value of the contribution and also because of the smearing of the transitions by inhomogeneities. So far, the critical exponents of the 3d XY model were not possible to be resolved in superconductors. However, deviations from the mean-field jump and the existence of anomalies associated with phase-fluctuation-driven transitions, i.e. vortex lattice melting, were observed~\cite{Roulin1996,Schilling1997,Bouquet2001}. This is an especially challenging experiment in our case since the splitting between the $Z_2$ and $U(1)$ phase transitions and the strength of superconducting fluctuations are very sensitive to small doping variation, as observed in this study and consistent with the 
discussion in \cite{Grinenko2021state,Bojesen2013,Bojesen2014a}. Here, we found that a few per cent changes in doping level narrow the splitting by about 50\%. The strongest splitting of about 1.8~K was observed for $S_{\rm NP}$ sample from Ref.~\cite{Grinenko2021state} with $x=0.77$, and it reduced to 1~K and lower for $x\approx0.75$.

The main result of this study is that in high-quality \BKFA{} samples, we observed two specific-heat anomalies.
One correlates with the onset of 
superconductivity while the other coincides with the spontaneous breaking of time-reversal symmetry, detected by the appearance of a spontaneous Nernst effect.
The breaking of the $Z_2$ symmetry above the superconducting transition temperature and its dependence on doping and external magnetic field
allowed earlier to establish that it is associated with the formation of a fermionic quadrupling order \cite{Grinenko2021state}. The current data provide calorimetric evidence for the existence of this novel phase in the \BKFA{} system at zero magnetic field. 
The second result of this study is the verification of the fermion quadrupling order at a different 
doping  from the one reported in \cite{Grinenko2021state}.

\section{Acknowledgments}

\begin{acknowledgments}
	E.B. was supported by the Swedish Research Council Grants 2016-06122, 2018-03659. F.C. acknowledges the European Union’s Horizon 2020 research and
innovation programme under the Marie Skłodowska-Curie grant agreement No 892728.  M.P. acknowledges Italian MUR projects: PRIN ‘HiBiSCUS’ Grant No. 201785KWLE. I.M. acknowledges the Carl Trygger foundation through grant number CTS 20:75. H.-H.K. acknowledges support by DFG grant SFB1143. We thank Ruidan Zhong for assistance in performing experiments.
	 
\end{acknowledgments}

\newpage
\bibliographystyle{naturemag}
\bibliography{weston-fixed}

\clearpage
\section{Supplementary information}
\renewcommand{\theequation}{S\arabic{equation}}
\renewcommand{\thefigure}{S\arabic{figure}}
\renewcommand{\thetable}{S\arabic{table}}
\setcounter{equation}{0}
\setcounter{figure}{0}
\setcounter{table}{0}

\subsection*{Samples}
Phase purity and crystalline quality of the plate-like Ba$_{1-x}$K$_x$Fe$_2$As$_2$ single crystals were examined by X-ray diffraction (XRD) and transmission electron microscopy (TEM). The K doping level $x$ of the single crystals was determined using the relation between the $c$-axis lattice parameter and the K doping obtained in previous studies \cite{Kihou2016}. The selected single-phase samples had a mass $\sim 0.1 - 1$~mg with a thickness $\sim 10 - 50$~$\mu$m and a surface area of several mm$^2$.

\subsection*{Experimental}
DC susceptibility measurements were performed using a commercial superconducting quantum interference device (SQUID) magnetometer from Quantum Design. The measurements of the specific heat using the thermal relaxation method were performed in a Quantum Design physical property measurement system (PPMS).  
The Nernst- and Seebeck-effect measurements were performed using a home-made sample holder for transport properties inserted in a Quantum Design physical property measurement system (PPMS) endowed with a $9$~T magnet.  The Seebeck coefficient ($S_{xx}$) is the ratio of the longitudinal electric field to the longitudinal thermal gradient applied to generate it. The Nernst coefficient ($S_{xy}$) is related to the transverse electric field produced by a longitudinal thermal gradient \cite{Behnia2016}. In order to create an in-plane thermal gradient on the bar-shaped samples, a resistive heater ($R = 2.7$~k$\Omega$) was connected on one side of the sample, while the other side was attached to a thermal mass. The temperature gradient was measured using a Chromel-Au-Chromel differential thermocouple, calibrated in magnetic field, attached to the sample with a thermal epoxy (Wakefield-Vette Delta Bond 152-KA). The Nernst and Seebeck signals were collected using two couples of electrodes (made of silver wires bonded to the sample with silver paint), aligned perpendicular to or along the thermal gradient direction, respectively. The magnetic field $B$ was applied in the out-of-plane direction. In order to separate the standard Nernst effect $S_{xy}$ from the spurious Seebeck component (caused by the eventual misalignment of the transverse contacts), the Nernst signal has been antisymmetrized by inverting the $B$ direction. The spontaneous Nernst signal, which is finite only in proximity to the superconducting transition, has been obtained by subtracting the Seebeck ($S_{xx}$) component from the $B$-symmetric part of the Nernst signal as described in Ref.~\cite{Grinenko2021state}. 

In the thermoelectric measurements, the temperature difference $\Delta T_{\rm sample}$ across the sample (measured by the thermocouple) did not exceed $3\%$ of the measurement temperature $T$ fixed by the thermal mass.

\begin{figure}
	\centering
	\includegraphics[width=0.95\linewidth]{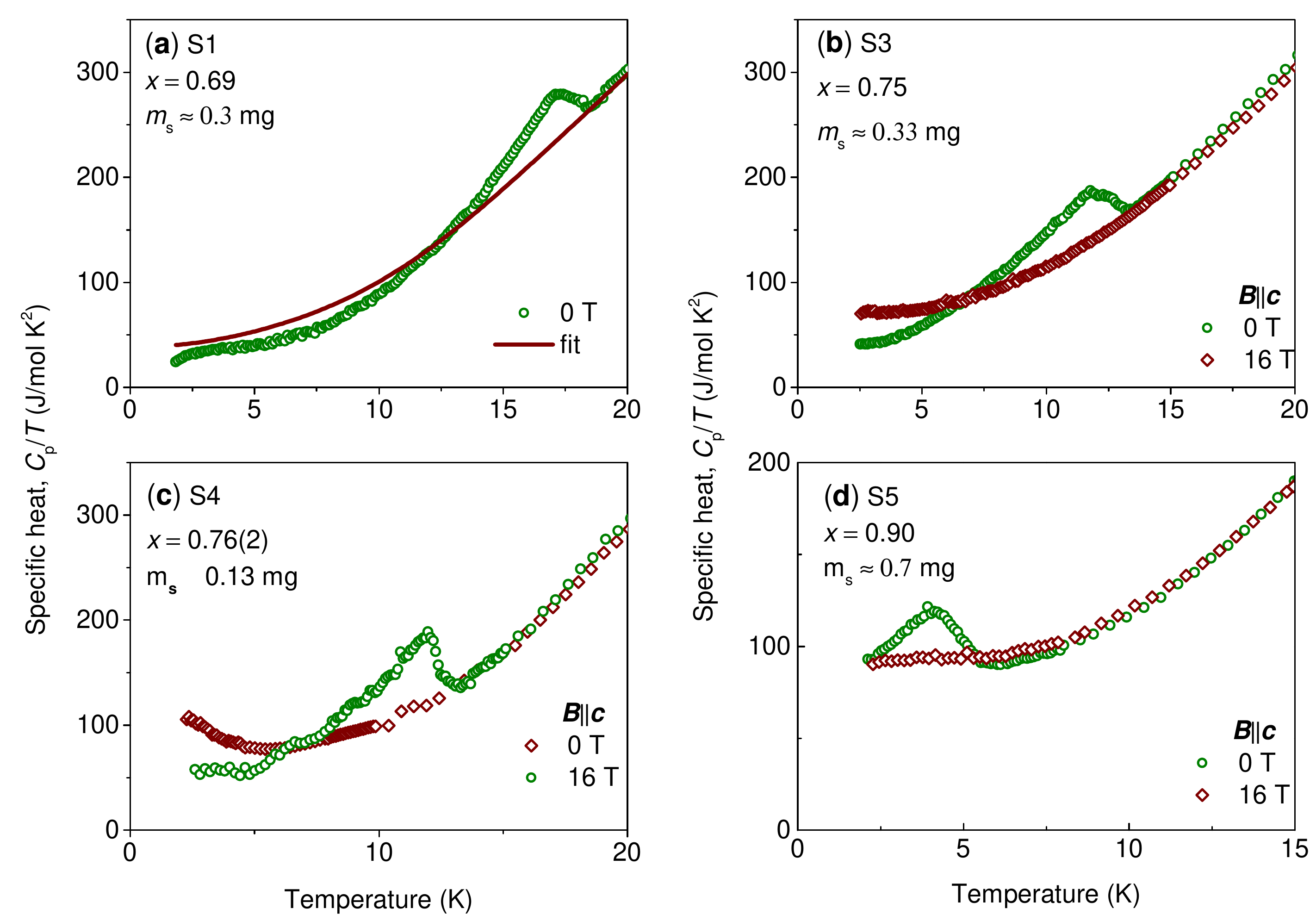}
	\caption{\textbf{Raw specific heat data}. Temperature dependence of the zero-field specific heat $ C_{\rm p}/T$ for the \BKFA{} samples $S1$, $S3$-$S5$ shown in the main text. The fitting curve in panel \textbf {(a)} is used to subtract the phonon background. The details of the fitting can be found in Ref.~\cite{Grinenko2020}. For samples $S3$-$S5$ shown in panels \textbf {(b-c)} the data measured in 16 T field applied along the crystallographic $c$-axis were used to obtain the phonon background. The results of the subtraction are shown in Fig.~2.}
	\label{figED1}
\end{figure}

\begin{figure}
	\centering
	\includegraphics[width=0.95\linewidth]{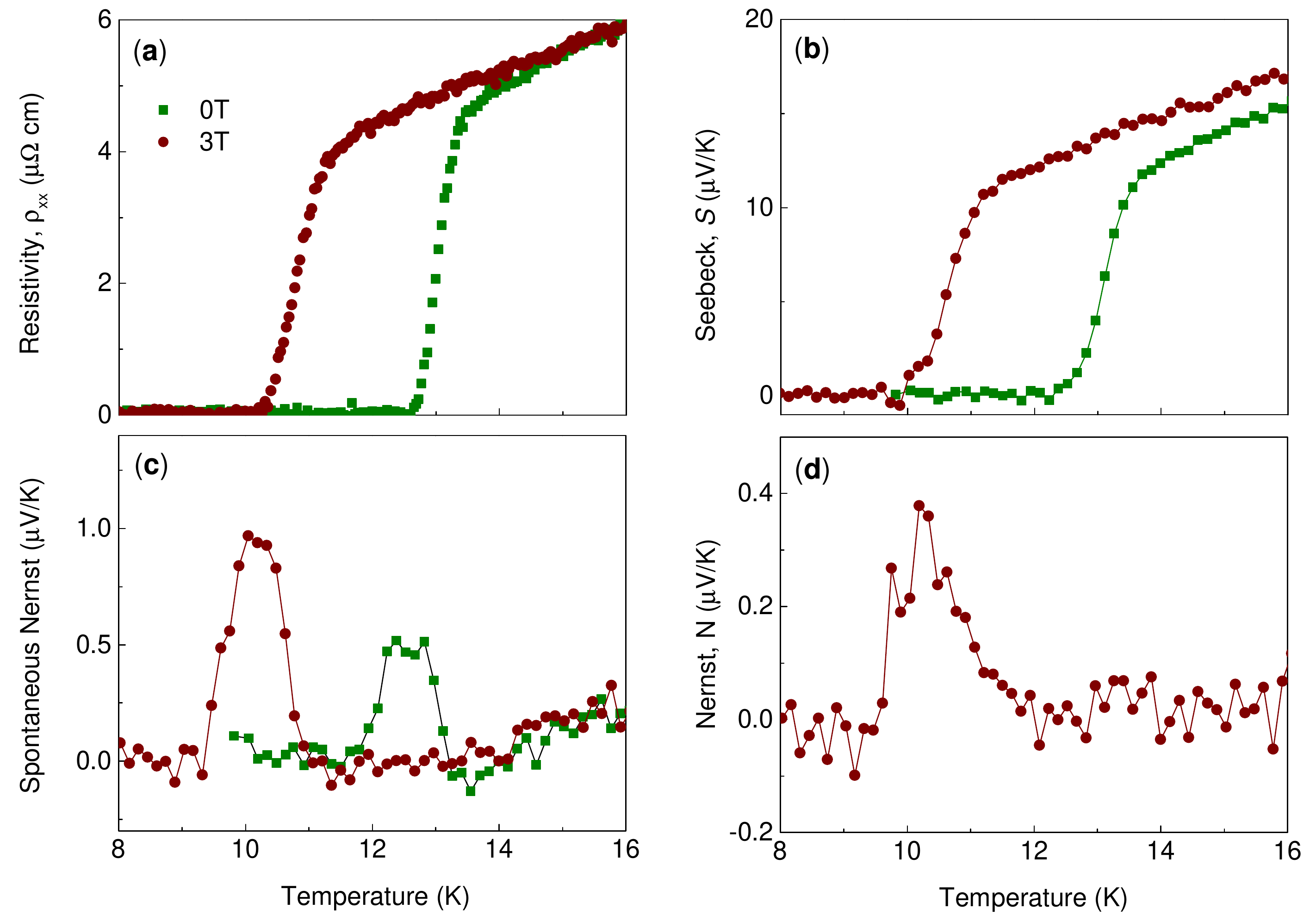}
	\caption{\textbf{Transport data for the sample $S2$} \textbf {(a)} Temperature dependence of the longitudinal electrical resistivity close to superconducting transition measured in zero and 3 T field applied along the crystallographic $c$-axis. \textbf{(b)} Temperature dependence of the Seebeck effect. The resistivity and Seebeck effect have a very similar temperature dependence.   
	\textbf{(c)} Temperature dependence of the spontaneous Nernst effect (even signal in a magnetic field). \textbf{(d)} Temperature dependence of the normal Nernst effect (odd signal in a magnetic field).}
	\label{figED2}
\end{figure}

\subsection{Details of Monte Carlo simulations}

The Monte Carlo (MC) simulations are performed by considering a three-dimensional cubic lattice of linear dimension $L$ and lattice spacing $h=1$. 
For the numerical calculations, we implement the Villain approximation~\cite{Villain1975} scheme, where the compactness of the phase difference is ensured by writing:  
\begin{equation*}
    e^{\beta \cos{ \left( \Delta_{\mu} \phi_{{\bf r}_i} \right)} }  \to \sum_{n=-\infty}^{\infty} e^{-\frac{\beta}{2} (\phi_{{\bf r}_{i+\mu}} - \phi_{{\bf r}_i}- 2\pi n)^2},
\end{equation*}
where $\Delta_{\mu} \phi_{{\bf r}_i} = \phi_{{\bf r}_{i+\mu}} -\phi_{{\bf r}_i} $ is the discrete phase difference between two nearest neighbours lattice sites ${\bf r}_{i+\mu}$ and ${\bf r}_i$ along $\mu=\hat{x}, \hat{y}, \hat{z}$, and $\beta=1/T$ is the inverse temperature.  The discrete Villain Hamiltonian for the model \eqref{f_2component} reads:
\begin{equation}
\begin{split}
        H_v[\phi_1, \phi_2, \beta] =  -\sum_{{\bf r}_i, \mu} \beta^{-1} \ln \left\{ \sum_{n_{1,\mu} n_{2,\mu}= - \infty}^{\infty} e^{-{\beta} S_{\mu}} \right\} ,
        \label{H_Vill}
\end{split}
\end{equation}
where
\begin{equation}
    S_{\mu} = \frac{1}{2} [u_{{\bf{r}}, \mu, 1}^2 + u_{{\bf{r}}, \mu, 2}^2] - \nu ( u_{{\bf{r}}, \mu, 1} \cdot u_{{\bf{r}}, \mu, 2}) + \eta_2 \cos[2(\phi_{{\bf{r}}, 1} -\phi_{{\bf{r}}, 2} )], 
\end{equation}
and $u_{{\bf{r}_i}, \mu, \alpha} = \Delta_{\mu} \phi_{{\bf{r}}, j} - 2\pi n_{{\bf{r}}, \mu, \alpha} $, with $\alpha=1,2$ label the two components. 
We performed MC simulations of the Villain Hamiltonian Eq.~\eqref{H_Vill}, locally updating the two phase fields $\phi_1, \phi_2 \in [0, 2\pi )$ by means of the Metropolis-Hastings algorithm. A single MC step here consists of the Metropolis sweeps of the whole lattice fields while, to speed-up the thermalization at lower temperatures, we implemented a parallel tempering algorithm. Typically, we propose one set of swap after 32 MC steps.
For the numerical simulations presented in this work, we performed a total of $2\times 10^5$ Monte Carlo steps, discarding the transient time occurring within the first $50000$ steps.

As discussed in~\cite{Grinenko2021state, Maccari2022effects}, we assessed the SC critical temperature $T_{\rm c}$ by computing the helicity-modulus sum $\Upsilon_+$. That is defined as the linear response of the system with respect to a twist of the two-component phases along a given direction $\mu$: 
\begin{equation}
\Upsilon^{\mu}_{+} =\frac{1}{L^3} \frac{\partial^2 F(\{\phi'_i\}) }{\partial \hat{\delta}_{\mu}^2}\Bigr|_{\hat{\delta}_{\mu}=0},
\label{Helicity_+}
\end{equation}
with: $\begin{pmatrix} \phi'_1(\mathbf{r}) \\  \phi'_2(\mathbf{r})  \end{pmatrix}
= \begin{pmatrix} \phi_1(\mathbf{r}) + \delta \cdot \mathbf{r}_{\mu} \\  \phi_2(\mathbf{r}) + \delta \cdot \mathbf{r}_{\mu} \end{pmatrix}$ and $\hat{\delta}_{\mu}=  \begin{pmatrix} \delta \cdot \mathbf{r}_{\mu} \\ \delta \cdot \mathbf{r}_{\mu} \end{pmatrix}$. In our MC simulations, the helcity-modulus sum has been computed along $\mu=\hat{x}$, so in what follows we mean $\Upsilon_+ \equiv\Upsilon^{x}_{+}$. 

The critical temperature $T_c$ is then extracted by taking the thermodynamic limit of the finite-size crossings of the quantity $L \Upsilon_+$, as shown in Fig.\ref{fig1}(a) for $\nu=0.6$ and $\eta_2=0.1$. 

On the other hand, we extracted the critical temperature $T_{\rm c}^{\rm Z2}$ by introducing a $Z_2$ Ising order parameter $m$, equal to $+1 $ or $-1$ according with the two possible sign of the ground-state phase difference $\phi_{1,2}=\pm \pi/2$. 
The $Z_2$ critical temperature is then determined by means of finite-size crossings of the Binder cumulant, associated with the Ising parameter $m$, extrapolated to the thermodynamic limit.
The Binder cumulant $U$

\begin{equation}
  U = \frac{\langle m^4 \rangle}{3\langle m^2 \rangle^2},
  \label{binder_cumulant}
\end{equation}
where $\langle \dots \rangle$ stays for the thermal average over the MC steps, is indeed expected to be a universal quantity at the critical point.
The finite-size crossing points of $U$ for $\nu=0.6$ and $\eta_2=0.1$ are shown in Fig.\ref{fig1}(b).

Finally, the specific heat $C_v$ shown in Fig.\ref{fig1}(c)-(d) is defined as:
\begin{equation}
    C_v= \frac{1}{T^2 L^3} \left[ \langle E^2\rangle - \langle E \rangle^2 \right], 
\end{equation}
where $E$ is the total energy of the system at a given temperature $T$.

The error bars of all the observables are estimated via a bootstrap resampling method. In the figures shown, when not visible, the estimated error bars are smaller than the symbol sizes.

\if


\end{document}